%% file: IWQoS.tex
\documentclass[sigconf,noacm]{acmart}
\AtBeginDocument{%
  }

\renewcommand\footnotetextcopyrightpermission[1]{} % remove the copyright footnote
\acmISBN{978-1-4503-XXXX-X/2018/06}

\usepackage{fix-cm}
\usepackage{amsmath,amsfonts} %amssymb
\usepackage{algorithmic}
\usepackage{graphicx}
\usepackage{textcomp}
\usepackage{comment}
\usepackage{xcolor}
\usepackage{algorithm}
\usepackage{subfigure}
\usepackage{multicol}
\usepackage{xspace}
\usepackage{multirow}
\usepackage{booktabs}
\usepackage{paralist}
\newcommand{\projname}{IDSS\xspace}
\usepackage[capitalise,nameinlink,noabbrev]{cleveref}
\usepackage{enumitem}
\usepackage{subcaption}

\usepackage[most]{tcolorbox}
\tcbuselibrary{skins,xparse,breakable}
\usepackage{amsthm}

\crefname{section}{\S}{\S{}}
\crefname{figure}{Fig.}{Fig.}
\crefformat{section}{\S#2#1#3}
\usepackage{multicol} 
\usepackage{booktabs}

\renewenvironment{itemize}{
 \begin{list}{\labelitemi}{
     \setlength{\topsep}{0.5ex}
     \setlength{\parsep}{0pt}
     \setlength{\partopsep}{0pt}
     \setlength{\itemsep}{2pt}
     \setlength{\itemindent}{10pt}
     \setlength{\leftmargin}{0pt}}
}{\end{list}}

\setlist[enumerate]{leftmargin=*}

\newcommand{\linebreakand}{%
  \end{@IEEEauthorhalign}
  \hfill\mbox{}\par      \mbox{}\hfill\begin{@IEEEauthorhalign}
}

\newcounter{designelemcnt}

\newcommand{\mypar}[1]{\paragraph*{\normalfont\textbf{#1.}}}
\crefname{section}{\S}{\S{}}
\crefname{figure}{Fig.}{Fig.}
\crefformat{section}{\S#2#1#3}
\usepackage{tikz}

\newcommand*\circled[1]{\tikz[baseline=(char.base)]{
            \node[shape=circle,fill,inner sep=1.4pt, scale=0.7] (char) {\textcolor{white}{#1}};}}

\begin{document}

\title[Intent-Driven Storage Systems: From Low-Level Tuning to High-Level Understanding]{Intent-Driven Storage Systems:\\From Low-Level Tuning to High-Level Understanding}

\author{Shai Bergman}
\email{shai.aviram.bergman@huawei.com}
\affiliation{%
  \institution{Huawei Zurich Research Center}
  \city{Zurich}
  \country{Switzerland}
}

\author{Won Wook Song}
\email{won.wook.song@huawei.com}
\affiliation{%
  \institution{Huawei Zurich Research Center}
  \city{Zurich}
  \country{Switzerland}
}

\author{Lukas Cavigelli}
\email{lukas.cavigelli@huawei.com}
\affiliation{%
  \institution{Huawei Zurich Research Center}
  \city{Zurich}
  \country{Switzerland}
}

\author{Konstantin Berestizshevsky}
\email{konstantin.berestizshevsky@huawei.com}
\affiliation{%
  \institution{Huawei Zurich Research Center}
  \city{Zurich}
  \country{Switzerland}
}

\author{Ke Zhou}
\email{zhke@hust.edu.cn}
\affiliation{%
  \institution{Huazhong University of Science and Technology}
  \city{Wuhan}
  \country{China}
}

\author{Ji Zhang}
\authornote{Corresponding author.}
\email{dr.jizhang@huawei.com}
\affiliation{%
  \institution{Huawei Zurich Research Center}
  \city{Zurich}
  \country{Switzerland}
}

\begin{abstract}

Existing storage systems lack visibility into workload intent, limiting their ability to adapt to the semantics of modern, large-scale data-intensive applications. This disconnect leads to brittle heuristics and fragmented, siloed optimizations.

To address these limitations, we propose Intent-Driven Storage Systems (\projname), a vision for a new paradigm where large language models (LLMs) infer workload and system intent from unstructured signals to guide adaptive and cross-layer parameter reconfiguration. \projname provides holistic reasoning for competing demands, synthesizing safe and efficient decisions within policy guardrails.

We present four design principles for integrating LLMs into storage control loops and propose a corresponding system architecture. Initial results on FileBench workloads show that \projname can improve IOPS by up to 2.45$\times$ by interpreting intent and generating actionable configurations for storage components such as caching and prefetching. These findings suggest that, when constrained by guardrails and embedded within structured workflows, LLMs can function as high-level semantic optimizers, bridging the gap between application goals and low-level system control. \projname points toward a future in which storage systems are increasingly adaptive, autonomous, and aligned with dynamic workload demands.

\end{abstract}

\keywords{Storage System, Configuration, Artificial Intelligence}

\maketitle
\pagestyle{plain}  
\sloppy
\input{content/1introduction}

\input{content/2motivation}

\input{content/3design}

\input{content/4feasibility}
\input{content/5related_work}

\input{content/6conclusion}

\bibliographystyle{ACM-Reference-Format}
\bibliography{IWQoS}
\end{document}

%% file: content/1introduction.tex
\section{Introduction}
\label{sec: intro}

Modern storage systems must meet increasingly diverse demands in a balanced manner: low latency for real-time applications, high throughput for analytics, and cost efficiency for archival workloads. This complexity is exacerbated by heterogeneous use cases such as AI training, online transaction processing (OLTP), and video streaming, which often coexist within a shared infrastructure~\cite{cao2020carver}.
Without coordinated management, these competing workloads contend for bandwidth, cache space, and IOPS, leading to unpredictable performance, inefficient resource utilization, and degraded quality of service across the system.

To manage performance, storage systems expose numerous tunable \emph{parameters} across components such as caching, quality of service (QoS), compression, and redundancy. A \emph{parameter} is an individual system knob (e.g. block size, cache eviction policy, or compression level) whose value affects runtime behavior. For instance, Linux's Ext4~\cite{ext4} offers about 60 parameters. A \emph{configuration} is a specific assignment of values to a set of such parameters, forming a complete system setup. For instance, in the XFS file system, the block size parameter can be set to values such as 512 bytes, 1K, 2K, or 4K; whereas a full configuration might look like \texttt{[blocksize=4KB, allocsize=64KB, inode size=512B]}. The set of all such valid combinations forms the system's \emph{parameter space}, a high-dimensional space that can contain as many as $10^{37}$ possible configurations for the Ext4 filesystem~\cite{cao2019practical,cao2018towards, cao2020carver}.

\noindent\textbf{The Challenge:} The vast size of the parameter space renders manual performance tuning error-prone and impractical~\cite{alkiswan2017}, while exhaustive automated approaches are computationally infeasible~\cite{cao2019practical}. Furthermore, this complexity poses significant challenges for system-wide coordination tasks, such as synchronizing caching strategies between clients and servers, making effective holistic optimization difficult to achieve in real-world deployments~\cite{zhang2009adding, macedo2020survey}.

Rule-based heuristics have traditionally been used to tune system parameters, but their effectiveness diminishes as workloads evolve beyond the assumptions of their designers~\cite{maheshwari1997}. Methods using machine learning, genetic algorithms, and simulated annealing~\cite{cheng2025,cao2018towards,akgun2023improving} remain limited with narrow configuration scopes, and insufficient consideration of workload semantics and client context~\cite{basak2016storage,soundararajan2008context}, constraining their generality and robustness.

Recent work has begun to explore the use of large language models (LLMs) for automating configuration tuning. For example, ELMo-Tune~\cite{thakkar2025elmo} uses LLMs to map natural language workload descriptions to low-level system configurations, demonstrating that LLMs can infer tuned parameters from unstructured input. However, these systems primarily focus on tuning configurations for individual servers or isolated workloads and do not provide the architectural mechanisms needed for coordinated optimization across multiple system layers or nodes. As a result, they fall short in adapting to dynamic, multi-tenant scenarios where decisions must account for shared resources and conflicting performance goals.

\noindent\textbf{Gaps in Current Solutions:} While the current approaches mark important progress, we argue that they remain inadequate in addressing three fundamental challenges:

\begin{enumerate}
    \item \emph{Intent blindness:} Systems are unable to deeply comprehend the specific intent, goal, and the semantic implications of the various workloads, resulting in one-size-fits-all, general policies~\cite{zhou2021learning}.
    \item \emph{General system complexity:} Modern storage systems involve numerous configurations over interdependent components~\cite{luttgau2018survey,cao2017performance}. Fragmented or layer-specific optimizations (e.g., garbage collection independent of caching) lack a holistic system perspective and overlook cross-layer dependencies, resulting in suboptimal resource utilization~\cite{zadok2015parametric,cao2020carver}. 
    \item \emph{Vendor lock-in:} Proprietary tools (e.g., Dell PowerMax QoS~\cite{dellpowermaxai2024}) cannot generalize across storage stacks and practical multivendor deployments.
\end{enumerate}

\noindent\textbf{We propose \projname}, an intent-driven \emph{storage agent} that autonomously configures, tunes, and orchestrates storage systems. \projname embodies a concrete design guided by a broader vision: enabling storage systems to adapt intelligently to diverse workload demands. It infers workload intent from unstructured context and translates it into coordinated, cross-layer configurations. It harmonizes client and server optimizations, for example, aligning client-side caching with server-side tiering, and reasons about the broader impact of such decisions across the stack. By bridging semantic gaps between administrators, applications, and heterogeneous hardware, \projname enables adaptive behavior that exceeds the limitations of traditional rule-based heuristics.

This paper presents the design principles of \projname{}, outlines the key challenges in realizing intent-driven storage, and proposes a system architecture to address them. We empirically validate essential LLM capabilities for \projname, including their ability to internalize unstructured domain knowledge, configure policies based on workload traces, and reason about cross-component interactions.

%% file: content/2motivation.tex
\section{Why LLMs for Storage Systems?}\label{BM}

Modern storage systems expose a vast configuration space where optimal performance depends on dynamic client workloads, storage server configuration, and resource interdependencies. Prior work by \citet{cao2017performance} demonstrates that a single configuration can yield a 40\% performance swing on one workload, but only 6\% on another, highlighting the sensitivity of system behavior to workload characteristics. Yet, traditional storage systems often lack insight into client workloads and their performance requirements, relying on naturally observable data such as client block requests. This narrow perspective restricts the system's ability to anticipate and adapt to changing workload demands. Static configurations and heuristic-based tuning methods frequently fall short in addressing these complexities, resulting in suboptimal throughput, increased latency, and inefficient resource utilization.

Human experts can partially mitigate this gap by manually correlating workload intent with system behavior and adjusting configurations accordingly. However, several limitations constrain the scalability and effectiveness of this approach.
First, no single expert possesses deep knowledge across all storage subsystems and workload types.
Second, human operators cannot continuously monitor dynamic workloads and system state at the granularity needed to support timely adaptation.
Third, manual tuning is often tailored to specific hardware and software configurations, limiting its ability to generalize across platforms or evolve with changing deployments.

LLMs exhibit strong zero-shot capabilities and cross-domain generalization~\cite{wang2022languagemodelarchitecturepretraining, brown2020language, raffel2020exploring}, enabling them to adapt seamlessly across diverse tasks such as natural language understanding, code generation, biomedical text analysis, and even multimodal reasoning. These models have been successfully applied in areas ranging from automated software debugging and healthcare diagnostics to financial forecasting and scientific literature mining, demonstrating their ability to transfer knowledge effectively between domains without task-specific fine-tuning~\cite{liu2023pre, qiu2020pre}. Importantly, LLMs excel at semantic reasoning, effectively parsing unstructured inputs (e.g., LogParser-LLM~\cite{zhong2024logParserLLM}) to determine optimized storage policies.

We therefore posit that \textbf{LLMs provide a compelling foundation for overcoming the limitations of manual and heuristic-based storage optimization}. Their capabilities span several dimensions critical to intent-aware system design:

%%NO TOUCHY! DONE - SHAI
\begin{enumerate}
    \item \emph{Goal-oriented reasoning:} LLMs can be prompted to infer workload-specific objectives~\cite{li2024towards}, such as prioritizing P99 latency for OLTP databases, and synthesize adaptive strategies that align with system constraints. Recent work also demonstrates their potential for use in resource planning and scheduling tasks~\cite{tang2025llm_schedule,abgaryan2024llmsschedule}. Moreover, while traditional optimization methods struggle with categorical parameters~\cite{eklund2019algorithms} like `deadline vs. cfq schedulers', an LLM-based system understands these choices contextually, reasoning about their trade-offs without artificial encoding schemes.
    \item \emph{Semantic bridging:} LLMs can close the information gap between clients and storage systems by representing both workload intent and system state in \emph{natural language}. Prior work has shown that LLMs can interpret client goals~\cite{anand2025intentbaseddesignoperation,jacobs2021hey} and parse system configurations and telemetry~\cite{ma2024llmparser}, enabling richer cross-layer understanding.
    \item \emph{Tool orchestration:} LLMs can automate system-wide configurations through function calling~\cite{chen2024enhancingfunctioncallingcapabilitiesllms} and adhere to predefined safety constraints, ensuring system stability.
    \item \emph{Generalized knowledge synthesis:} Trained on decades of research papers, documentation, and logs, LLMs internalize best practices across storage architectures and vendors~\cite{ji2024adaptinglargelanguagemodels}. Moreover, LLMs can swiftly expand their knowledge by leveraging external data sources via retrieval augmented generation (RAG)~\cite{lewis2020rag}, thereby unlocking more information for better decision-making~\cite {lee-etal-2024-planrag}.
\end{enumerate}

By serving as storage agents, LLMs offer a unifying layer that integrates storage systems, client behavior, workload semantics, and domain knowledge into a coherent decision-making framework. In this role, they act as a ``system of systems'', coordinating insights and actions across otherwise siloed components.

Recent advances in enterprise deployment of local AI agents~\cite{azure-ai-agent2024} suggest that LLM-based storage agents are increasingly feasible in practice. However, their integration introduces new challenges, ranging from safety and performance to abstraction boundaries, which we outline and address through a set of design principles in the following section.

%% file: content/3design.tex
\section{Design}
\label{sec:design}

\subsection{Principles for Intent-Driven Storage Servers}

We identify four key challenges in designing intent-driven storage systems that leverage LLMs, and propose corresponding design principles to address them. These principles demonstrate how LLMs can help tackle previously intractable problems, such as cross-layer optimization and vendor-specific policy translation, while maintaining relatively low engineering overhead. At the same time, they incorporate safeguards to mitigate risks such as configuration hallucinations, using structured guardrails and controlled execution boundaries.

\noindent\textbf{P1: Autonomous, context-aware adaptation}\newline
\emph{Challenge:} Traditional storage systems rely on static configurations or heuristic rules that do not generalize across workloads or adapt to changing conditions. This often leads to suboptimal performance and resource over-provisioning~\cite{zadok2015parametric}.\newline
\noindent\emph{Principle:} Storage systems should infer workload requirements from high-level application semantics (e.g., identifying an OLAP database implies prioritizing low-latency random reads) and dynamically adapt policies as workloads and system conditions evolve.\newline
\noindent\emph{LLM-driven opportunity:}
\begin{enumerate}
    \item LLMs can translate unstructured context, such as workload names, descriptions, or telemetry, into actionable system policies~\cite{anand2025intentbaseddesignoperation}. For example, given a video streaming workload characterized by sequential access patterns, an LLM can recommend bandwidth reservation and local pre-buffering of video segments~\cite{4783016}. Notably, such decisions can incorporate unstructured performance insight without requiring rigid APIs or deep integration efforts.
    \item LLMs can autonomously adjust configurations using new research, hardware specifications, and API documentation, without requiring manual retraining via RAG~\cite{fu2024autoraghpautomaticonlinehyperparameter}.
\end{enumerate}

\noindent\textbf{P2: Holistic, system-wide optimization}\newline
\emph{Challenge:} Storage systems are often optimized in isolation across layers (e.g., caching, garbage collection) and components (e.g., clients and servers). This siloed approach leads to systemic inefficiencies such as redundant data movement, misaligned caching policies, and uncoordinated resource usage.\newline
\noindent\emph{Principle:} 
Storage systems should coordinate configuration decisions across interdependent layers and distributed components. Effective optimization requires reasoning about cross-layer dependencies and system-wide telemetry, including second-order effects introduced by a single policy change.\newline
\noindent\emph{LLM-driven opportunity:}
\begin{enumerate}
    \item LLMs can leverage domain expertise to adjust interdependent parameters. For instance, correlate deduplication intensity with SSD wear-out models, throttling redundant writes when drive health metrics degrade~\cite{6232379}.
    \item LLMs can interpret workload intent to harmonize configurations across components~\cite{Dzeparoska2023llmBasedPolicy}. This includes disabling redundant server-side caching when client-side hit rates exceed 90\%, or aligning client prefetching with server-tiering policies to reduce I/O contention.
\end{enumerate}

\noindent\textbf{P3: Guarded autonomy through structured control flow}\newline 
\emph{Challenge:} LLM-driven configuration, like human expert tuning, carries the risk of producing unsafe or suboptimal decisions, potentially violating performance objectives. \newline
\noindent\emph{Principle:} 
To ensure safe and predictable behavior, LLM-generated actions must be governed by a structured control flow that decomposes decisions into modular, auditable steps. At each stage, proposed actions are validated against deterministic safety checks before execution.
In addition, systems should version and persist previously successful configurations, enabling rollback in the event of unexpected performance regressions. A/B testing mechanisms could additionally be employed to evaluate new configurations under controlled conditions before full deployment, providing a safety net even in the presence of guardrails.
\newline
\noindent\emph{LLM-driven opportunity:}
\begin{enumerate}
\item LLM reasoning can be modularized across discrete operational stages, enabling contextual focus and targeted validation while reducing exposure to long-context errors.
\item Safety safeguards can incorporate hallucination mitigation techniques drawn from LLM code generation research, such as cross-checking against retrieved documentation~\cite{eghbali2024dehallucinator,zhang2025llmhallucinationspracticalcode,anonymous2024hallucheck,zhang2024knowhalu}.
\item When uncertainty remains high, the system can trigger clarification prompts to augment the LLM's input with richer context, as demonstrated by ClarifyGPT~\cite{mu2024clarifygpt}.
\end{enumerate}

\noindent\textbf{P4: Vendor-neutral policy abstraction}\newline
\emph{Challenge:} Storage systems face vendor lock-in due to incompatible configuration formats and APIs, requiring manual policy translation across platforms~\cite{Razavian2013analsis}.\newline
\noindent\emph{Principle:} 
To enable portability and extensibility, storage systems should decouple policy logic from vendor-specific interfaces. This requires adopting an expressive intermediate representation (IR) that abstracts away vendor-specific differences while still allowing platform-specific optimizations. This mirrors compiler architecture, where an ISA-agnostic IR supports code portability without sacrificing backend specialization.\newline
\noindent\emph{LLM-driven opportunity:}
\begin{enumerate}
    \item LLMs can leverage expressive natural language as a vendor-agnostic IR, bypassing low-level syntax barriers.
    \item LLMs equipped with RAG can query vendor documentation to automatically translate high-level policies into platform-native configuration commands.
\end{enumerate}

\begin{figure*}[t]
  \centering
  \includegraphics[width=\textwidth]{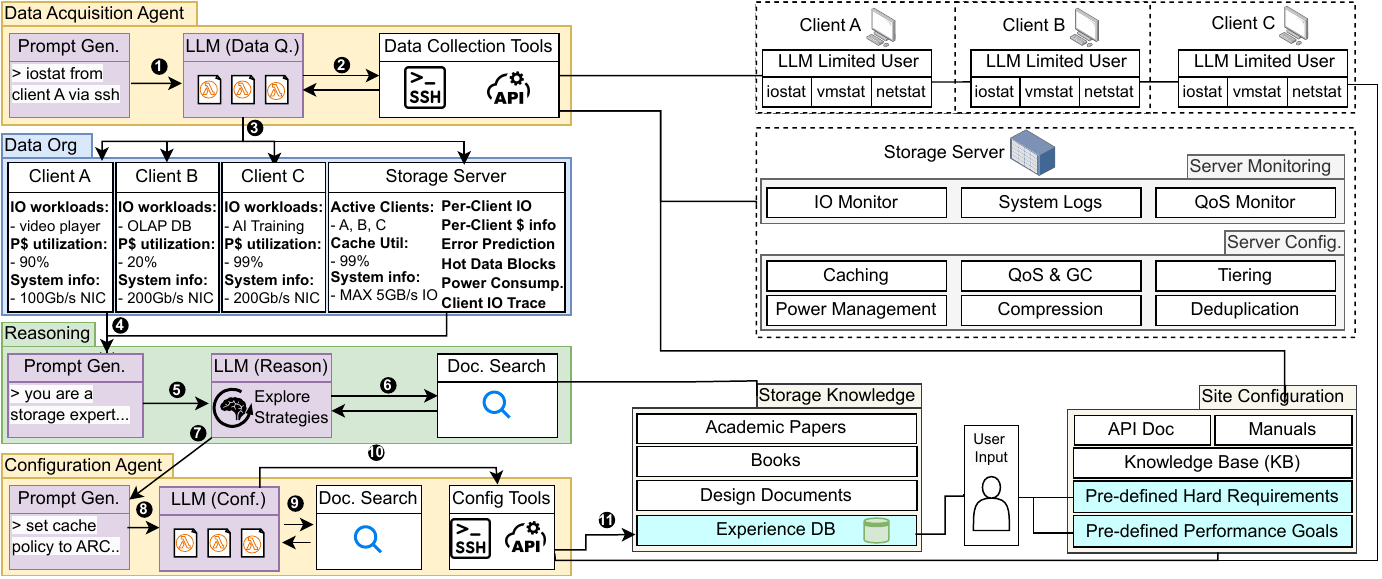}
  \Description{}
  % \vspace{-3mm}
  \caption{Overview of the \projname framework for system-wide parameter optimization.}
  \label{fig:overview}
  % \vspace{-1mm}
\end{figure*}

\subsection{Design Overview}

\cref{fig:overview} presents the proposed architecture of \projname, which integrates intent-driven reasoning powered by LLMs, into the configuration and control of storage systems. The design follows the four principles outlined in the previous section and is structured into four interconnected phases: Data Acquisition, Data Organization, Reasoning, and Configuration, each responsible for transforming input signals into actionable, validated system policies.

\noindent\emph{Data Acquisition Agent:} The agent initiates the workflow by dynamically generating prompts to collect telemetry from clients and the storage server, such as I/O statistics, client and server cache utilization, and site configurations with the user's hard requirements~\circled{1}. 
The LLM translates these data acquisition prompts into executable commands, employing secure remote access protocols to collect client-side data and vendor-specific APIs for server-side operations~\circled{2}. The LLM's data acquisition API calls are performed with respect to the vendor's API library available at the moment, supporting design principle P4, Vendor agnostic policy abstraction.

\noindent\emph{Data Organization:} 
Raw metrics are organized into a structured representation, retaining only predefined information and performance-critical metrics~\circled{3}. Additionally, the structured data explicitly links client workload to its current system state, establishing a clear foundation for reasoning.  Importantly, the aggregated system-wide data organization structure materializes our design principle P2, Holistic, system-wide optimization.

\noindent\emph{Reasoning:}
The module begins by creating a prompt that combines a predefined high-level system objective~\cite{thakkar2025elmo} with structured system information~\circled{4}. The high-level objective leads the LLM's reasoning mechanism to analyze the current workloads and clients' data, identifying performance targets specific to each workload, such as prioritizing low latency for OLAP databases while ensuring stable bandwidth for video streaming~\circled{5}. It also enforces administrator-defined constraints and predefined goals, such as ``minimize the promotion of data to SSD tiers to preserve endurance''. The aforementioned data acquisition, organization, and reasoning are based on design principle P1, Context-aware adaptation.

To generate configuration strategies for the clients and the storage system, the LLM queries the Storage Knowledge repository~\circled{6}, which includes the system's design documents, research papers, and an experience database, to identify context-aware optimizations. For example, when recognizing Client B’s OLAP workload, the LLM infers the need for lower tail latency and adjusts the I/O scheduler to prioritize its requests. This bridges the semantic gap that persists with rule-based systems. Additionally, the LLM can assess how multiple clients interact to affect overall system performance, reasoning how changes to a single client impact the overall performance goals. Following design principle P3, the LLM can reduce hallucinations by engaging in a feedback loop with the Storage Knowledge Repository, issuing clarification requests to augment its prompt and refine its reasoning~\cite{mu2024clarifygpt}.

The configuration strategy produced by the LLM reasoning module is handed off to the configuration agent~\circled{7}, which translates it into executable, vendor-agnostic actions targeting both clients and the storage server. The agent then invokes the LLM~\circled{8}, which leverages the Operational Knowledge repository~\circled{9} to generate platform-specific commands, such as SSH and API calls. Finally, the agent executes these actions via structured function calls~\circled{10}, applying the configuration safely across the system. The LLM's function calls are performed with respect to the available system's software/hardware API library, supporting our design principle P4, Vendor agnostic policy abstraction.

\projname updates the Experience DB with new configurations, performance statistics, and conclusions learned from past experiences~\circled{11}. The Experience DB can be initialized with several stable configurations to serve as a solid fall-back plan during the operation or good starting points for further optimizations. However, its most important role is to provide context for high-quality reasoning.

\mypar{LLM Configuration for Safety and Consistency}
The effectiveness of \projname relies critically on how its LLM components are configured during generation. To ensure reliable and factual responses across different agents, the system must constrain the model's randomness and control its output behavior. For example, limiting the range of likely next-word predictions helps avoid unsupported or overly speculative responses, while still allowing for some flexibility to avoid rigid or repetitive errors.

Additional safeguards include narrowing the output length and enforcing context-sensitive stopping conditions, which prevent the model from generating off-topic or verbose outputs. Recent work also suggests dynamically adjusting the model's response variability based on confidence or uncertainty, improving the balance between precision and adaptability~\cite{zhang2025llm,eghbali2024dehallucinator,tonmoy2024comprehensive}. Collectively, these generation-time controls form an essential layer of safety and consistency in LLM-driven storage systems, ensuring that each reasoning step remains interpretable, grounded, and aligned with operational goals.

%% file: content/4feasibility.tex
\begin{figure*}
  \centering
  \includegraphics[width=0.8\textwidth]{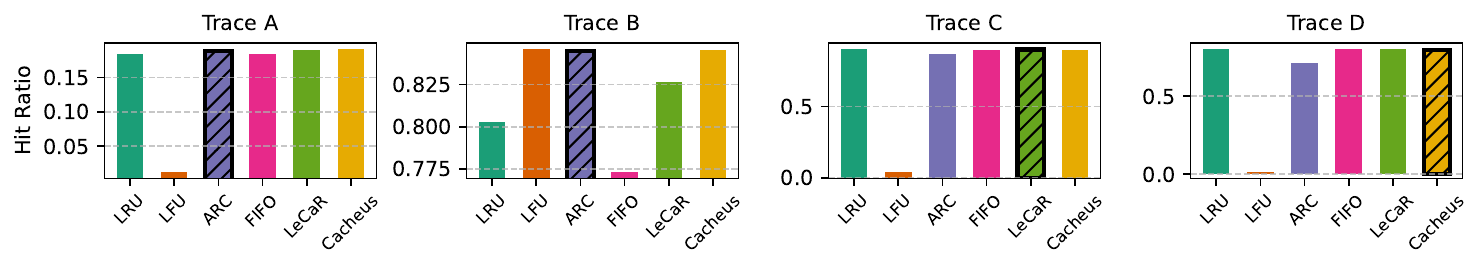}

  \caption{Synthetic traces. Hatched bar shows the policy selected by the LLM.}

  \label{fig:synthetic}
\end{figure*}

\begin{figure*}
  \centering
  \includegraphics[width=\textwidth]{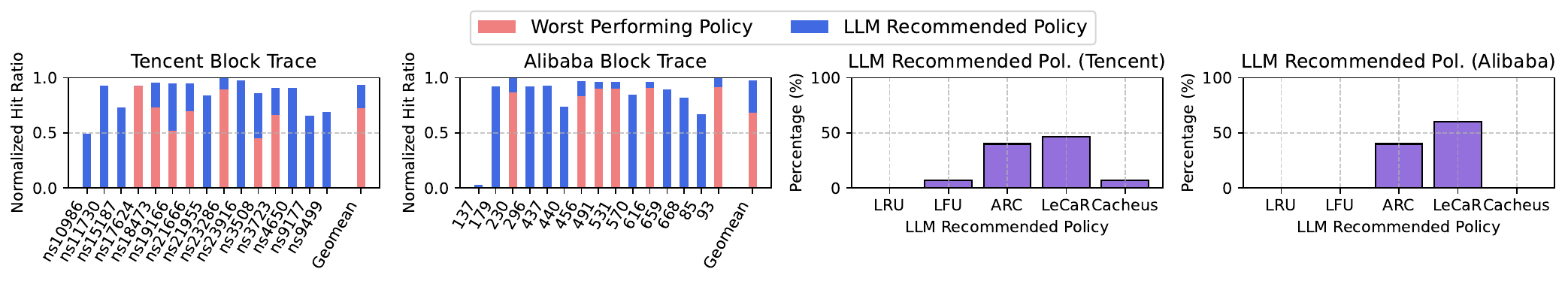}

  \caption{Cache policy performance and distribution for real-world block I/O traces.}
  \label{fig:real}
\end{figure*}
\section{Experimental Insights}
\label{sec:feasible}

To validate \projname's vision of autonomous LLM-driven storage agents, we evaluate three fundamental capabilities of LLMs: (1) semantic reasoning to infer workload intent, (2) operational automation for translating intent into actionable steps, and (3) adaptive decision-making for data-driven configuration optimization. 
Finally, we use the FileBench benchmark~\cite{Tarasov2016FilebenchAF} as a macro-level evaluation to assess how these components interact when deployed end-to-end.
Our experiments are conducted using OpenAI's API~\cite{openai2023api}.

\mypar{Semantic reasoning}
We evaluate LLMs' ability to infer storage requirements from unstructured telemetry data using sample outputs from \texttt{iotop}~\cite{iotop}. Specifically, we collect these outputs for three representative workloads: an OLTP database (MySQL~\cite{mysql}), media streaming (FFmpeg~\cite{ffmpeg}), and AI checkpointing (PyTorch~\cite{rojas2021studycheckpointinglargescale}). To enable richer analysis, we also supply detailed system and filesystem metadata, including the filesystem type (e.g., ext4), block size, and journal configuration.

The LLM successfully performed workload classification, extracted I/O requirements, and generated tailored configuration suggestions. 
In a coordination scenario involving multiple concurrent workloads, we provided telemetry for a video streaming workload sustaining 100\,MB/s reads and an AI checkpointing workload with bursty writes peaking at 1.5\,GB/s. The LLM recommended reserving 1.2\,GB/s of bandwidth for checkpointing and capping streaming reads at 300\,MB/s to minimize contention. It further suggested enabling 256\,KB read-ahead exclusively for the streaming workload, avoiding cache pollution from OLTP's random accesses.

These results demonstrate the LLM's capacity to fulfill \emph{P1: Autonomous, context-aware adaptation} and \emph{P2: Holistic, system-wide optimization}. The model exhibited reasoning grounded in workload semantics, generating workload-specific policies rather than defaulting to one-size-fits-all configurations.

\mypar{From intent to execution}
To evaluate the feasibility of intent-driven operational automation with LLMs, we tested the model's ability to generate executable scripts and OS-specific (e.g., Linux) commands from natural language context while parsing vendor API documentation. The LLM translated high-level objectives, such as ``create a QoS class for video streaming with a 500\,MB/s bandwidth cap" into vendor-specific API instructions. Inputs combined natural language prompts with API manuals to mirror real-world deployment scenarios where administrators must reconcile intent with platform constraints.

We further introduced strict operational guardrails by prefacing prompts with system limitations, for example, ``NIC bandwidth capped at 100\,MB/s". The LLM internalized these constraints during its reasoning process, iteratively validating proposals against the provided guidelines.

These results demonstrate that LLMs can align with \emph{P3: Guarded autonomy through structured control flow} and \emph{P4: Vendor-neutral policy abstraction}. As observed in prior work~\cite{wies2023subtaskdecompositionenableslearning}, decomposing tasks into discrete stages minimizes hallucinations by bounding the LLM's reasoning scope. However, success depends critically on integrating RAG with up-to-date vendor documentation.

\mypar{Adaptive decision-making from raw data}
A core challenge is deriving actionable insights from low-level telemetry to complement workloads' intent, such as block access traces. To evaluate whether LLMs can reason over raw, unstructured data series, we conduct experiments to test the LLM's ability to infer suitable cache replacement policies from partial traces. This task requires pattern recognition, temporal reasoning, and domain knowledge~\cite{song-blaze}. We generate four synthetic block traces:
\begin{itemize}
    \item A: 1K preloaded blocks followed by 5K random accesses.
    \item B: 80\% of accesses to 100  frequently accessed blocks.
    \item C: Cyclic reuse of a contiguous 1K blocks.
    \item D: 5 epochs of contiguous 2K-block active set.
\end{itemize}
For each trace, we provided the LLM with the first 400 requests and tasked it with selecting a policy from LRU, LFU, FIFO, ARC, LeCaR, and Cacheus. We then evaluated all the traces for all policies using libcachesim~\cite{yang2020-workload}, configured with a cache size of 0.1\% of the working set.

\Cref{fig:synthetic} compares cache hit rates for the evaluated traces and replacement policies, with highlighted bars indicating the LLM's recommendations. The results demonstrate that the LLM consistently selected policies achieving near-optimal performance, within 2\% of the best-performing policy, while avoiding choices exhibiting significant hit ratio degradation.

We extend our evaluation to 30 real-world block I/O traces from Alibaba~\cite{alibaba} and Tencent~\cite{zhang2020osca}. \Cref{fig:real} shows the normalized hit rates of LLM-recommended policies relative to the best-performing alternative (excluding FIFO, which underperformed across all cases). The LLM's choices achieved a geometric mean of 97\% of the best policy's hit rate, outperforming the worst policy (excluding FIFO) by 1.45$\times$.

\Cref{fig:real} also illustrates the LLM's policy selections across workloads, shown as a histogram. The distribution reveals that the model selects different caching strategies depending on the workload, validating its context-aware reasoning and adaptive behavior, rather than relying on a fixed default.

These results demonstrate the potential of LLMs to reason over raw telemetry and trace data, enabling \emph{P2: Holistic, system-wide optimization} even under limited sampling. 

\begin{figure}[t]
  \centering
  \includegraphics[width=\columnwidth]{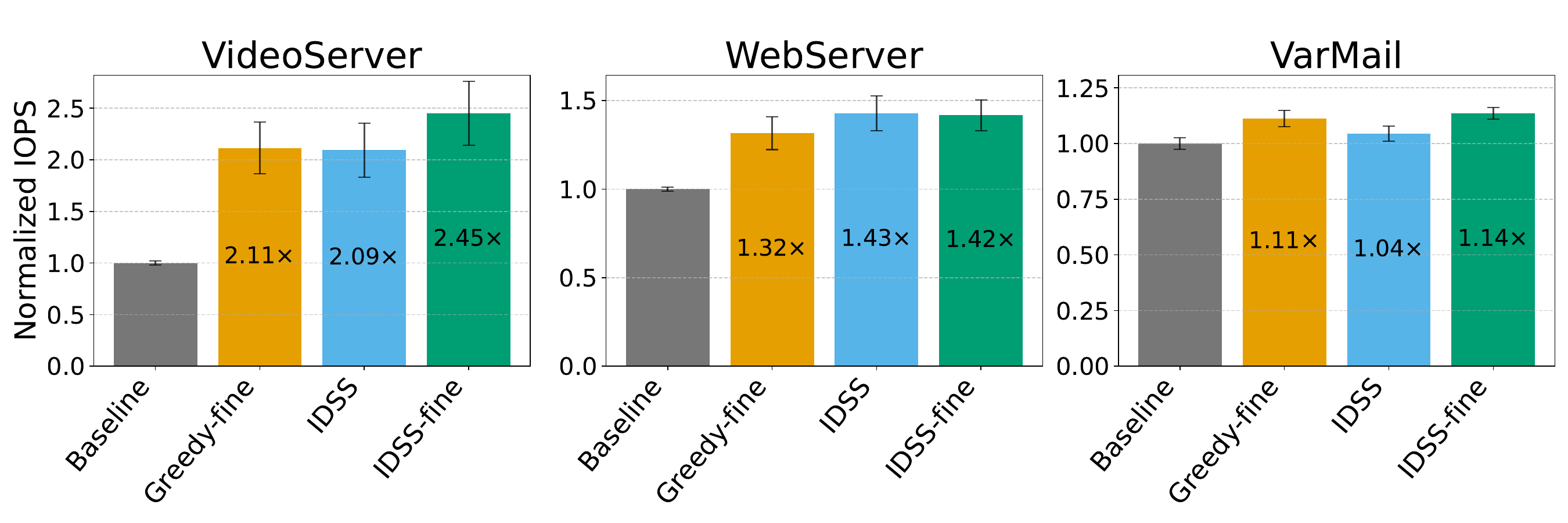}
  
  \caption{Comparison of different methods of storage system parameter tuning on different workloads. 
  }
  \label{fig:workloads}
\end{figure}

\mypar{Filebench Evaluation}
To assess \projname's effectiveness across realistic application mixes, we evaluate the LLM's decision-making capabilities using contextual inputs tailored to representative workload profiles from the FileBench benchmark suite~\cite{Tarasov2016FilebenchAF}. These include VideoServer, WebServer, and VarMail, each exhibiting distinct I/O patterns: extensive sequential reads, read-dominant access with occasional log writes, and frequent small file creation, deletion, and \texttt{fsync} operations, respectively.

Experiments were conducted on a system with 2$\times$64-core ARM Kunpeng-920 CPUs, 256\,GiB of DRAM, and a 4$\times$ SAS SSD array. To ensure that the SSDs themselves do not limit the performance, we monitored the SSD utilization and confirmed sufficient headroom. All benchmark processes were consistently pinned to the same CPU cores and NUMA nodes across runs to eliminate variability unrelated to file system configuration.

Our evaluation centers on tuning key parameters of the Ext4 file system, including journal settings (e.g., write barriers), block size, and I/O scheduler, and compares the performance across four different configurations:
\begin{itemize}
    \item \emph{Baseline} - the default Ext4 settings.
    \item \emph{Greedy-Fine} - a Carver-inspired baseline~\cite{cao2020carver} that incrementally selects influential parameters and performs black-box optimization using LLM queries.
    \item \emph{\projname{}} - using generic, non-workload-specific context.
    \item \emph{\projname-Fine} - using workload-specific context derived from the ``Client A/B/C'' descriptors in the Data Organization phase.
\end{itemize}

Each benchmark was run five times per configuration. The median IOPS results are shown in \cref{fig:workloads}. Compared to the Baseline, Greedy-Fine achieves improvements of $2.11\times$ (VideoServer), $1.32\times$ (WebServer), and $1.11\times$ (VarMail). \projname-Fine further increases performance to $2.45\times$, $1.42\times$, and $1.14\times$ for the same workloads. Even without workload-specific tuning, \projname delivers notable gains of $2.09\times$, $1.43\times$, and $1.04\times$, respectively.

While Greedy-Fine demonstrates competitive performance in select cases, it lacks consistency across workloads. In contrast, \projname-Fine consistently outperforms all baselines by leveraging workload-specific context to generate more nuanced configurations (e.g., increase the read-ahead size for the VideoServer). Notably, although \projname underperforms Greedy-Fine slightly in the VarMail workload, the fine-tuned variant (\projname-Fine) recovers the performance gap and surpasses both baselines.

These results demonstrate \projname's potential for generalizable, intent-driven optimization across diverse storage workloads. As more components from our broader vision (\cref{fig:overview}) are integrated, such as an experience database, external knowledge sources, and structured telemetry, further performance and robustness improvements are anticipated.

%% file: content/5related_work.tex
\section{Related Work}

Rule-based systems rely on predefined heuristics to guide configuration decisions~\cite{rule_based_1}. While simple and interpretable, such systems lack the flexibility to adapt to dynamic workloads or unforeseen runtime conditions, as their logic is hardcoded and context-agnostic~\cite{10.1145/3126908.3126932}. This rigidity often leads to degraded performance in heterogeneous or evolving environments. 

To overcome these limitations, a range of optimization-based approaches have been proposed. These include simulated annealing and genetic algorithms~\cite{cao2018towards}, supervised~\cite{6339587,ibrahim2021machine} and unsupervised learning~\cite{aken2017automatic,eml}, and deep reinforcement learning~\cite{zhangJis_paper_about_RL_for_cloud_DB}. These methods can generalize better than rule-based systems, but typically require significant task-specific model tuning and feature engineering~\cite{mukherjee2023towards}. Carver~\cite{cao2020carver}, for example, reduces the configuration space of storage servers by selecting a small set of influential parameters using conditional importance metrics. While effective, its reliance on sampled data and iterative evaluation limits its ability to capture complex parameter interactions. Techniques like Carver may serve as complementary components within \projname, enhancing decision quality during LLM-assisted reasoning.

Recent works have begun to explore LLMs for configuration tuning. NetLLM~\cite{wu2024netLLM} applies LLMs to networking tasks such as adaptive bitrate streaming, while ELMo-Tune~\cite{thakkar2024can,thakkar2025elmo} targets LSM key-value stores by mapping workload descriptions to parameter sets. Dzeparoska \textit{et al.}~\cite{Dzeparoska2023llmBasedPolicy} propose an LLM-driven control loop that interprets natural language intent and generates corresponding policies, an approach aligned with our vision. While these efforts showcase the versatility of LLMs in domain-specific
tuning, they remain focused on isolated tasks.

%% file: content/6conclusion.tex
\section{Conclusion and Future Work}
\projname is a storage agent design that leverages the transformative potential of LLMs in bridging semantic gaps across storage systems, enabling autonomous, context-aware optimization through its design principles. Our experiments validate LLMs’ ability to infer workload intent, synthesize vendor-agnostic policies, and perform cross-layer decisions and safe operational automation. In future work, we wish to realize this vision, extend it with self-reflection to help guide future storage system design, and investigate the additional challenges posed, such as inference latencies.